\documentclass{aa}
\usepackage{graphicx}
\usepackage{txfonts}
\begin{document}
   \title{First observation of Jupiter by {\it XMM-Newton}}

   \author{G. Branduardi-Raymont
          \inst{1}
          \and
          R. F. Elsner
          \inst{2}
          \and
          G. R. Gladstone
          \inst{3}
          \and
          G. Ramsay
          \inst{1}
          \and
          P. Rodriguez
          \inst{4}
          \and
          R. Soria
          \inst{1}
          \and
          J. H. Waite, Jr
          \inst{5} 
                    }

   \offprints{G. Branduardi-Raymont
              \email{gbr@mssl.ucl.ac.uk}}

   \institute{Mullard Space Science Laboratory, University College London,
              Holmbury St Mary, Dorking, Surrey RH5 6NT, UK
    \and
              NASA Marshall Space Flight Center, SD50, 
              Huntsville AL 35812, USA
    \and
              Southwest Research Institute, San Antonio, Texas 78228, USA
    \and
              XMM-Newton SOC, Apartado 50727, Villafranca, 28080 Madrid,
              Spain
    \and
              University of Michigan, Space Research Building,
              2455 Hayward, Ann Arbor, Michigan 48109, USA
              }

   \date{Received 22 April 2004; accepted 23 May 2004}

   \abstract{
We present the first X-ray observation of Jupiter by {\it XMM-Newton}.
Images taken with the EPIC cameras show prominent emission, essentially all
confined to the 0.2$-$2.0 keV band, from the planet's auroral spots; their
spectra can be modelled with a combination of unresolved emission lines of
highly ionised oxygen (OVII and OVIII), and a pseudo-continuum which may
also be due to the superposition of many weak lines. A 2.8$\sigma$
enhancement in the RGS spectrum at 21$-$22 \AA\ ($\sim$0.57 keV) 
is consistent with an 
OVII identification. Our spectral analysis supports the hypothesis that 
Jupiter's auroral emissions originate from the capture and acceleration 
of solar wind ions in the planet's magnetosphere, followed by X-ray 
production by charge exchange. The X-ray flux of the North spot is modulated 
at Jupiter's rotation period. We do not detect evidence for the $\sim$45 min 
X-ray oscillations observed by {\it Chandra} more than two years earlier.
Emission from the equatorial regions of the planet's disk is also observed. 
Its spectrum is consistent with that of scattered solar X-rays.  

\keywords{planets and satellites: general -- planets and satellites 
individual: Jupiter -- X-rays: general}
   }

   \maketitle

\section{Introduction}
Jovian auroral X-ray emissions were first observed with the {\it Einstein
Observatory} (Metzger et al. \cite{Metzger}) and were extensively studied with
{\it ROSAT} (e.g. Waite et al. \cite{Waite94}, Gladstone et al. 
\cite{Gladstone98}),
which provided limited spectral information but fairly extensive imaging
data. The emissions have been explained as the result of charge exchange
and excitation of energetic ($>$1 MeV per nucleon) S and O ions (Cravens 
et al. \cite{Cravens}, \cite{Cravens03}). The ions were first thought 
to originate in Io's volcanos, 
and to precipitate from a region of Jupiter's magnetosphere just outside the 
Io Plasma Torus (IPT) at about 8 -- 12 Jovian radii (Mauk et al. \cite{Mauk}). 
This idea had to be reconsidered, though, because of more recent (December 
2000) {\it Chandra} observations: they have shown that most of Jupiter's 
northern auroral X-rays come from a hot spot located poleward of the latitudes
connected to the inner magnetosphere, pointing to a particle population in
the outer magnetosphere beyond 30 Jovian radii (Gladstone et al. 
\cite{Gladstone02}).
The magnetic mapping of the hot spot to such large distances presents 
serious difficulties regarding the source of the precipitating particles, 
since at $>$30 Jovian radii there are insufficient S and O ions to account 
for the hot spot emissions. Lack of correlation between the X-ray emission 
morphology and the surface magnetic field strength also suggests
that some process other than energetic ion precipitation from the 
inner magnetosphere is responsible for the bulk of the auroral X-rays.
One possibility is high-latitude reconnection of the planetary and solar 
wind magnetic fields, with the subsequent entry of the highly ionised (but 
low energy) heavy ion component (such as highly charged O, Fe, etc.) 
of the solar wind. The ions could then be accelerated to MeV energies by the 
currents present in the outer magnetosphere, thereby producing a spectrum rich 
in emission features much like comets. Indeed {\it Chandra} ACIS-S spectra 
(Elsner et al. 2004, in preparation) indicate a role for
energetic O and S ion precipitation as a source of the X-ray aurora: they
show a strong OVIII line at 653 eV, which puts the charge exchange 
X-ray production, pioneered by Horanyi et al. (\cite{Horanyi}) and Cravens 
et al. (\cite{Cravens}), on firm observational ground.

Another intriguing aspect of Jupiter's hot spot X-ray emissions observed by 
{\it Chandra} in December 2000 are $\sim$45 min pulsations, similar to those 
reported for high-latitude radio and energetic electron bursts observed by the 
{\it Ulysses} spacecraft during a fly-by a decade before (MacDowall et al. 
\cite{MacDowall}, McKibben et al. \cite{McKibben}). 
However, no comparable periodicity was seen in solar wind or interplanetary
magnetic field measurements made by {\it Cassini} upstream of Jupiter at the 
time of the {\it Chandra} observations in 2000, indicating that the pulsations
must arise from processes internal to the Jovian magnetosphere,

In addition, there is the open question of the source of Jupiter's 
low-latitude X-rays. Maurellis et al. (\cite{Maurellis}) have recently 
suggested that much of the equatorial X-ray emission can be understood 
by the scattering of solar X-ray photons by Jupiter's atmosphere. However, 
Waite et al. (\cite{Waite97}) argued that the 
correspondence between the low-latitude surface magnetic field and the 
observed longitudinal asymmetries in the X-ray emission observed by {\it ROSAT}
can best be explained by energetic S and O ion precipitation from the inner 
radiation belts. 

We set out to use the unparalleled combination of grasp and energy resolution 
of the {\it XMM-Newton} payload to investigate some of the un-answered 
questions
concerning Jupiter's X-ray emission. Reported here is the initial 110 ks 
observation performed in April 2003: its primary aim was that of verifying 
the feasibility of pointing at an optically bright target such as Jupiter.
A more recent, 
longer (245 ks) {\it XMM-Newton} observation carried out in November 2003 
will be presented in a future publication.

\section{Observations}
{\it XMM-Newton} (Jansen et al. \cite{Jansen}) observed Jupiter for 110 ks
between 2003, April 28, 16:00 and April 29, 22:00. The EPIC-MOS (Turner et
al. \cite{turner}) and pn (Str\"{u}der et al. \cite{struder}) cameras
(with a field of view of 30\arcmin\ diameter)
were operated in Full Frame and Large Window mode respectively,
and the RGS instrument (den Herder et al. \cite{denHerder}) in Spectroscopy.
The filter wheel of the OM telescope (Mason et al. \cite{Mason}) was kept in
the BLOCKED position because the optical brightness of Jupiter is above the
safe limit for the instrument, so no OM data were collected; also to minimise
the risk of optical contamination the EPIC cameras were used with the thick
filter. For a solar-type spectrum this filter is able to suppress efficiently
the optical contamination for point sources up to magnitude 0 (pn) and 1 (MOS);
this is valid for the worst case, of the brightest pixel within the
core of the Point Spread Function (PSF), and for Full Frame operational modes.
During the April 2003 {\it XMM-Newton} observation Jupiter had a surface
brightness of 5.5 mag/arcsec$^2$ (from the JPL HORIZONS Ephemeris
Generator). Thus we do not expect optical contamination in the EPIC-MOS
cameras (1.1\arcsec\ $\times$ 1.1\arcsec\ pixels) nor in the pn
(4.1\arcsec\ $\times$ 4.1\arcsec\ pixels). This is confirmed by the lack
of evidence for a steep rise in the spectra towards the lowest 
energies (see Fig.s~\ref{fig6},~\ref{fig7} and ~\ref{fig8}).

Jupiter's motion on the sky (11\arcsec/hr) required eight pointing 
trims during the long observation in order to avoid worsening the RGS spectral 
resolution; the planet's path was very close to the RGS dispersion direction, 
so that the RGS spectra of Jupiter's two poles, well separated spatially
(the planet's disk diameter was 38\arcsec\ during the observations),
did not overlap.

\begin{figure}
   \centering
   \includegraphics[width=6cm]{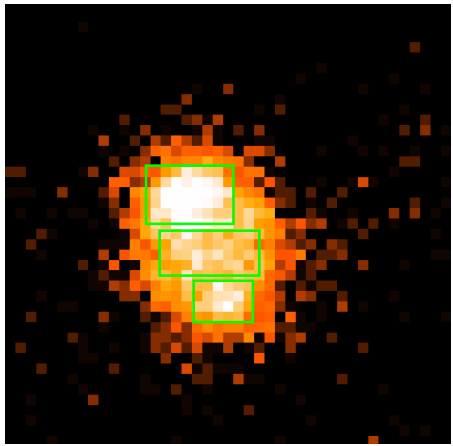}
      \caption{Jupiter's image (2\arcmin\ side) obtained combining EPIC-pn, 
MOS1 and MOS2 data; North is to the top and East to
the left. Superposed are the rectangular regions used in the extraction
of auroral and disk lightcurves and spectra. }
         \label{fig1}
   \end{figure}

The data were analysed with the {\it XMM-Newton} Science Analysis Software 
(SAS) v. 5.4 (see SAS User's Guide at 
http:// xmm.vilspa.esa.es/external/xmm\_user\_support/documentation). Photons 
collected during the nine pointings along Jupiter's path were referred to the 
centre of the planet's disk. An image of the planet, obtained by combining 
EPIC-pn, MOS1 and MOS2 data, is shown in Fig.~\ref{fig1}. A field source was 
occulted by Jupiter's equatorial region at the beginning of the observation: 
because of its low countrate (one fifth of the planet's disk) and of 
the short time ($\sim$1 hr) for eclipse ingress and egress, the source
is expected to produce no contamination on Jupiter's equatorial emission.

\begin{figure}
   \centering
   \includegraphics[width=8cm]{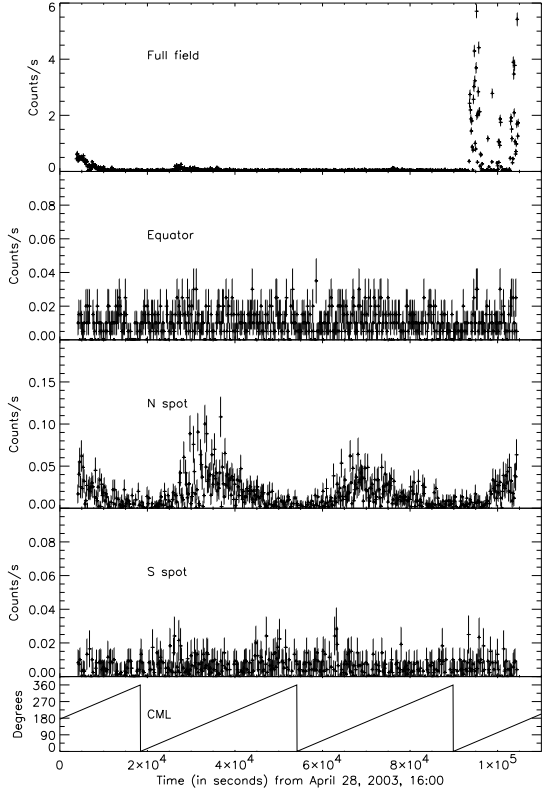}
      \caption{EPIC-pn lightcurves from the observation of Jupiter (Top panel:
full camera field at energies $>$ 10 keV, in 100 s bins, showing times of 
high particle background; middle panels: 0.2$-$2.0 keV, 200 s bins; 
bottom panel: Central Meridian Longitude; see text for details).}
         \label{fig2}
   \end{figure}

The top panel of Fig.~\ref{fig2} shows the EPIC-pn lightcurve (100 s bins) for
the full camera at energies $>$ 10 keV, which gives 
a good indication of the temporal behaviour of the particle background.
Beginning and end of the observation are affected by higher background
levels; excluding such times, a low-background period of 80 ks (2003, April 
28, 19:07 $-$ April 29, 17:20) is available and further analysis was 
restricted to this.

\section{EPIC timing analysis}

When extracting lightcurves (and spectra; see Sect. 4) for the auroral regions
we have taken into account the fact that the {\it XMM-Newton} telescope
PSF will 'mix' the auroral events with some from the
planet's disk. In order to establish the amount of such contamination
we convolved {\it Chandra} ACIS images (PSF of about 0.8\arcsec\
Half Energy Width, or HEW) with the {\it XMM-Newton} telescope PSF
($\sim$15\arcsec\ HEW), and estimated the
percentages of auroral, disk and off-planet diffuse background events falling
in rectangular regions of different sizes positioned at the auroral spots.
As extraction regions we selected
larger rectangles (24\arcsec\ x 16\arcsec\ for the North spot and
16\arcsec\ x 12\arcsec\ for the South) in preference to smaller ones
because they allow us to include a larger number of auroral photons, with
only a relatively small increase in the contributions of disk and
background events. A rectangular box (28\arcsec\ $\times$ 13\arcsec) was also 
used to extract Jupiter's equatorial lightcurve and spectrum.
The extraction regions are shown in Fig.~\ref{fig1}, superposed on the 
combined EPIC-pn, MOS1 and MOS2 image of Jupiter: we expect that
in the boxes over the North and South spots 71 and 66\% of the events are
of auroral origin, 27 and 31\% of disk origin, and 2 and 3\% are from the
off-planet diffuse background, respectively. The latter contribution is small 
enough that it was neglected when extracting the auroral lightcurves (and 
spectra). A $\sim$2\%
contribution of particle background (estimated from an out-of-field-of-view 
area of the MOS1 camera) was also neglected. 

EPIC-pn lightcurves for Jupiter's equatorial region and the planet's North 
and South auroral spots (the latter two after subtraction of the appropriate 
contribution from Jupiter's disk) are shown in the middle panels of 
Fig.~\ref{fig2} (in 200 s bins and in the energy range 0.2$-$2.0 keV,
where essentially all the X-ray emission is detected). 

The planet's $\sim$10 hr rotation period is clearly seen in the lightcurve of
the North spot (third panel from top in Fig.~\ref{fig2}), but is apparently 
not visible for the (weaker) South spot (bottom but one panel) and the 
equatorial region (second panel from top).
We have further investigated the possible presence of periodic
variability in the X-ray emissions of Jupiter by generating amplitude
spectra for the lightcurves (in 240 s bins) of the events originating in 
the equatorial and the two auroral regions individually. 

\begin{figure}
   \centering
   \includegraphics[width=6cm,angle=90]{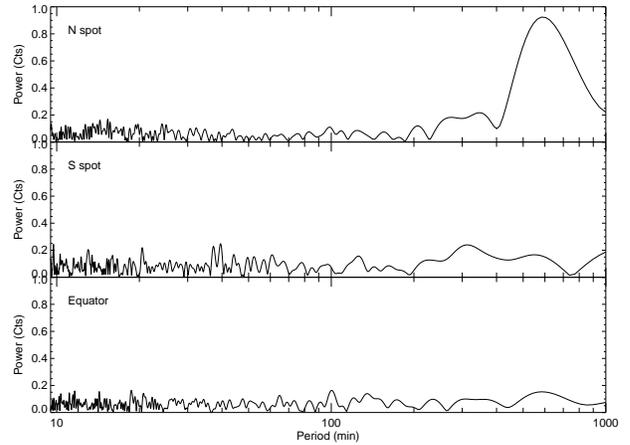}
      \caption{EPIC-pn amplitude spectra for Jupiter's North and South spots
and equatorial region.}
         \label{fig3}
   \end{figure}

\begin{figure}
   \centering
   \includegraphics[width=8cm]{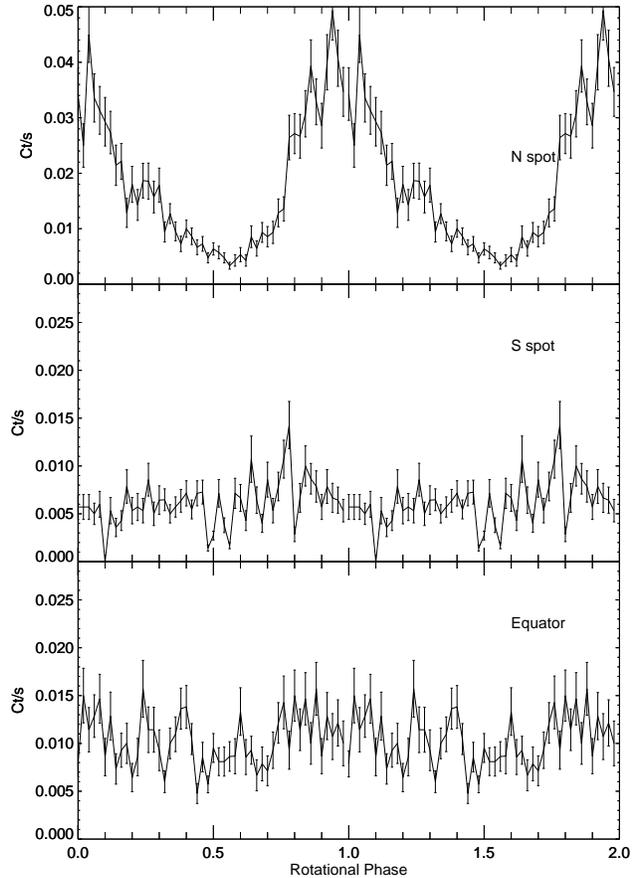}
      \caption{EPIC-pn lightcurves folded on Jupiter's 10 hr rotation
              period.}
         \label{fig4}
   \end{figure}

The amplitude spectra (Fig.~\ref{fig3}) clearly show power at Jupiter's 
10 hr rotation period in the North spot, but not in the South spot nor in 
the equatorial data. This is confirmed by inspection of the lightcurves 
folded on the 10 hr period (Fig.~\ref{fig4}). 

The bottom panel in Fig.~\ref{fig2} shows the Central Meridian Longitude
(CML, i.e. the system III longitude of the meridian facing the observer)
for the duration of the {\it XMM-Newton} observation. From the plot it is
clear that the North spot is brightest around longitude 180\degr, which 
is similar to what {\it Chandra} found (Gladstone et al. \cite{Gladstone02});
for the equatorial region and the South spot the distribution in longitude 
is more or less uniform (again similar to {\it Chandra}, which revealed a band,
rather than a hot spot, of emission near the South pole).

A search for periodic behaviour at shorter timescales (e.g.
the $\sim$45 min oscillations observed by {\it Chandra} in December 2000) shows
that there are individual peaks in the power spectrum below 200 min (see 
Fig.~\ref{fig3}), but that they are not significant: we proved this by taking 
the N spot lightcurve, randomly populating the data bins (of 240 s each)
and generating the events amplitude spectra. This was repeated 100 times,
with the result that no peak in the observed data is greater than the general
noise in the simulated data. A separate power spectrum analysis of the section 
of the lightcurve between 2$\times 10^4$ and 4$\times 10^4$ s into the 
observation (which shows a 'burst'-type behaviour) also does not produce any 
significant peak. Amplitude analysis of the EPIC-MOS data produces results 
consistent with those from the pn. 

We conclude that the $\sim$45 min oscillations observed by {\it Chandra} 
in December 2000 were not present, or were below the level of detectability, 
during the {\it XMM-Newton} observation of April 2003. 
It is worth pointing out that subjecting the {\it XMM-Newton} data
to the same analysis carried out on the {\it Chandra} data shows a very
similar level of amplitude noise at periods shorter than $\sim$30 min.

\section{EPIC spectral analysis}

\begin{figure}
   \centering
   \includegraphics[width=6cm]{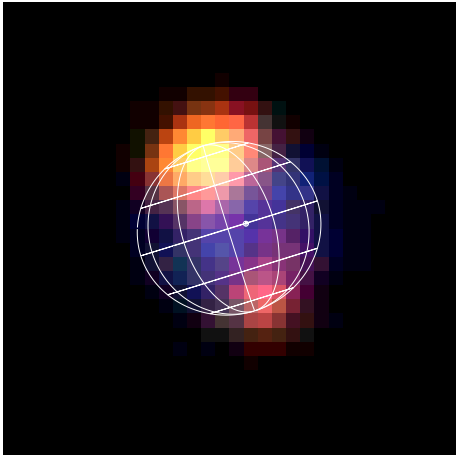}
      \caption{Smoothed EPIC-pn image of Jupiter (2.9\arcsec\ pixels);
North is to the top and East to the left; colour code: red - 0.2$-$0.5 keV, green - 0.5$-$0.7 keV,
blue - 0.7$-$2.0 keV. A graticule showing Jupiter's orientation with 30\degr\
intervals in latitude and longitude is overlaid. The circular mark indicates 
the sub-solar point; the sub-Earth point is at the centre of the graticule.}
         \label{fig5}
   \end{figure}

Fig.~\ref{fig5} shows a smoothed EPIC-pn colour image of Jupiter (where red 
corresponds to 0.2$-$0.5 keV, green to 0.5$-$0.7 keV and blue to 
0.7$-$2.0 keV) which clearly demonstrates that the equatorial disk emission 
is harder than that of the auroral spots. 

EPIC-pn, MOS1 and 2 spectra for Jupiter's auroral spots and disk were
extracted using the SAS task {\tt xmmselect}, selecting only good quality
(FLAG = 0) events. The appropriate percentage of disk
events (see Sect. 3) was then subtracted from the
auroral spectra, after normalising to the different areas of the extraction
regions. No background was subtracted from the disk spectrum. The resulting 
spectra were combined to produce integrated EPIC
spectra for the North and South spots and the equatorial disk region; 
finally the spectra were binned so as to have a S/N ratio
greater than 3$\sigma$ (North spot) or at least 20 counts (South spot and
equator) in each bin: in this way the $\chi^2$ minimisation technique 
is applicable in the fits. The spectra were fitted, using {\tt 
XSPEC} v. 11.3.0, in the energy range 0.2$-$2 keV, which contains essentially 
all the signal from Jupiter. Response matrices and auxiliary response files
for the three EPIC cameras and each of Jupiter's three regions
were built using the SAS tasks {\tt rmfgen}
and {\tt arfgen}. Since most or all of the flux comes from inside 
the extraction boxes and there is no source flux from outside the boxes 
falling into them, we adopted the point source option in {\tt arfgen}, because 
this will produce a better approximation of the real flux. The response
matrices for the three EPIC cameras were then combined (following Page et 
al. \cite{Page}) before being convolved with the spectral models in 
the fitting procedure for each region of the planet.

\subsection{North and South auroral spots}

A {\tt mekal} coronal plasma model and 
models comprising a number of Gaussian emission lines, with and without a 
continuum component, were tried. 

The best {\tt mekal} fits to the auroral regions are obtained for a temperature
of $\sim$0.2 keV with solar abundances, but the values of $\chi^2$/d.o.f. are
very high (106/44 and 59/25 for the North and South spots respectively): 
the overall shape of the model is a poor match to the observed spectrum and 
the sharp peak at $\sim$0.6 keV. 
For both auroral regions the best fit ($\chi^2$/d.o.f. = 46/40 and 20/21
for North and South spot respectively) is found with a power law continuum
and five Gaussian emission lines corresponding to CVI Ly$\alpha$ 
(0.37 keV), the OVII triplet (0.57 keV), OVIII Ly$\alpha$ (0.65 keV),
a combination of higher order OVII transitions (effective energy 0.69 keV)
and a blend of higher order OVIII 
lines (effective energy 0.80 keV). The energies of the lines were fixed 
in the fits and their intrinsic widths were set to zero (the latter are 
found to be below the EPIC resolution, or consistent with zero, if they are 
allowed to vary in the fits).

\begin{figure}
   \centering
   \includegraphics[width=5.5cm,angle=-90]{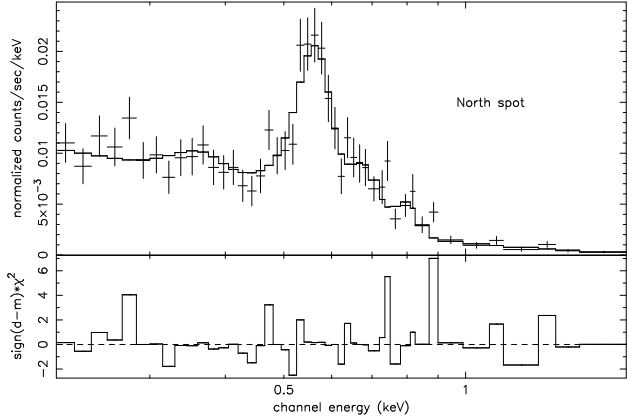}
      \caption{Data and best fit for the EPIC-pn spectrum of Jupiter's 
North auroral spot (see text for details).}
         \label{fig6}
   \end{figure}

\begin{figure}
   \centering
   \includegraphics[width=5.5cm,angle=-90]{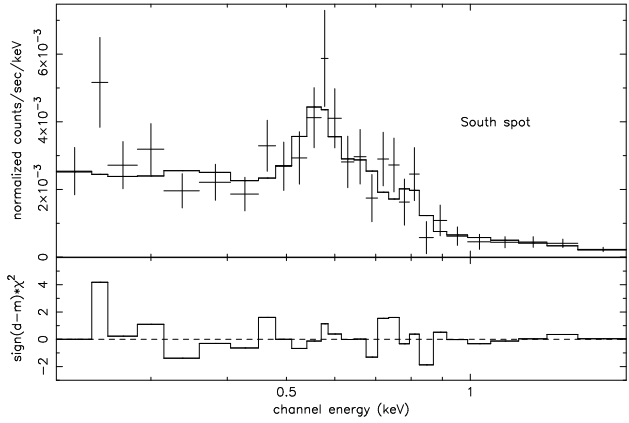}
      \caption{Data and best fit for the EPIC-pn spectrum of Jupiter's 
South auroral spot (see text for details).}
         \label{fig7}
   \end{figure}

The relevant best fit parameters (with errors; these are quoted at 90\%
confidence throughout the paper) are given in Table 1;
data, best fit model and the contribution of each bin to the $\chi^2$
are shown in Fig.~\ref{fig6} and Fig.~\ref{fig7} for the North and the
South spot respectively. The measured 0.3$-$2.0 keV flux is 5.6
$\times$ 10$^{\rm -14}$ erg cm$^{\rm -2}$ s$^{\rm -1}$ for the
North spot, and 2.1 $\times$ 10$^{\rm -14}$ erg cm$^{\rm -2}$ s$^{\rm -1}$ 
for the South.

\begin{figure}
   \centering
   \includegraphics[width=5.5cm,angle=-90]{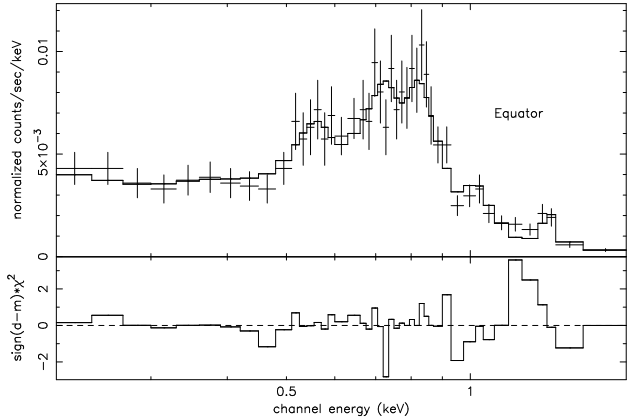}
      \caption{Data and best fit for the EPIC-pn spectrum of Jupiter's 
equatorial region (see text for details).}
         \label{fig8}
   \end{figure}

\begin{table}
      \caption{Best fit parameters for the 0.2$-$2~keV spectra of Jupiter's
auroral spots (errors are at 90\% confidence).}
         \label{fits}
$
         \begin{array}{lccccc}
            \hline\hline
            \noalign{\smallskip}
 & & \multicolumn{2}{c}{North~spot} & \multicolumn{2}{c}{South~spot} \\
 & & \Gamma^a & Norm^b & \Gamma^a
& Norm^b \\
            \noalign{\smallskip}
            \hline
            \noalign{\smallskip}
 Power~law &  & 3.23^{\rm +0.20}_{\rm -0.25} & 4.59^{\rm +0.66}_{\rm -0.69}
& 2.68^{\rm +0.53}_{\rm -0.38} & 3.02^{\rm +0.63}_{\rm -0.71}\\
            \noalign{\smallskip}
            \hline
            \noalign{\smallskip}
Emiss.~lines & Energy^{c} & Flux^{d} & 
EW^{e} & Flux^{d} & EW^{e} \\
            \noalign{\smallskip}
            \hline
            \noalign{\smallskip}
CVI~Ly\alpha~  & 0.37    & 14.28^{\rm +5.18}_{\rm -5.13} &
130^{\rm +47}_{\rm -47} &  3.89^{\rm +3.57}_{\rm -3.78} &
88^{\rm +81}_{\rm -86} \\
            \noalign{\smallskip}
OVII~triplet~   & 0.57    & 17.69^{\rm +2.54}_{\rm -2.33} &
638^{\rm +92}_{\rm -84} &  4.30^{\rm +1.71}_{\rm -1.59} & 
305^{\rm +121}_{\rm -113} \\
            \noalign{\smallskip}
OVIII~Ly\alpha & 0.65   &  1.54^{\rm +1.45}_{\rm -1.29} &
85^{\rm +56}_{\rm -71} &  0.93^{\rm +1.26}_{\rm -0.93} &
110^{\rm +149}_{\rm -110} \\
            \noalign{\smallskip}
OVII~higher~    & 0.69    &  3.42^{\rm +1.02}_{\rm -0.96} &
228^{\rm +68}_{\rm -64} &  0.97^{\rm +0.83}_{\rm -0.97} & 
94^{\rm +80}_{\rm -94} \\
            \noalign{\smallskip}
OVIII~higher~    & 0.80   &  1.68^{\rm +0.64}_{\rm -0.60} &
180^{\rm +69}_{\rm -64} &  0.82^{\rm +0.55}_{\rm -0.57} & 
144^{\rm +97}_{\rm -100} \\
            \noalign{\smallskip}
            \hline\hline
            \end{array}
$\\

$^{a}$ Photon index\\
$^{b}$ Power law normalisation at 1 keV in units of
10$^{\rm -6}$ ph cm$^{\rm -2}$ s$^{\rm -1}$ keV$^{\rm -1}$\\
$^{c}$ Energy of the emission features in keV (fixed in the 
fits)\\
$^{d}$ Total flux in the line in units of
10$^{\rm -6}$ ph cm$^{\rm -2}$ s$^{\rm -1}$\\
$^{e}$ Line equivalent width in eV\\
\end{table}

We explored the possibility that CV (0.31 keV) rather than CVI emission 
may be present at the low energy end of the spectra by letting the energy 
of the line free in the fits: for the North spot we find a best fit energy 
of 0.36$\pm$0.02 keV ($\chi^2$/d.o.f. = 45/39); fixing the line at 0.31 keV
gives a much worse fit with $\chi^2$/d.o.f. = 53/40. For the South spot, 
the best fit energy is 0.33$^{\rm +0.02}_{\rm -0.03}$ ($\chi^2$/d.o.f. = 
17/20), marginally consistent with CV emission; the rather poor 
statistical quality of the data, though, makes line discrimination uncertain.

With a view at establishing the origin (solar wind or inner magnetosphere)
of the ions responsible for the X-ray production in the auroral spots,
we checked whether the carbon emission could be
replaced by sulphur lines: we considered lines (or blends) at 0.32 keV (SXI),
0.34 keV (SXII) and 0.35 keV (SXIII) (some of the strongest in the atomic
spectral tables of Podobedova et al. \cite{Podo}; see also Elsner et al. 2004,
in preparation). While the energy resolution of the EPIC CCDs ($\sim$100 eV
FWHM below 1 keV), combined with the low countrate, make it very hard
to distinguish between lines that are
so close in energy, the best fits obtained by letting the line energy free
(see above) appear to indicate, at least for the North spot, preference
for a higher energy, i.e for the CVI~Ly$\alpha$ line (0.37 keV).

We find similar line intensities to those listed in Table 1, but slightly 
worse fits, if we substitute a bremsstrahlung continuum for the power law 
($\chi^2$/d.o.f. = 48/40 and 
25/21 for North and South spot respectively). Replacing the continuum component
with a combination of emission lines also produces worse quality fits.

\subsection{Equatorial region}

As suggested by Fig.~\ref{fig5}, we find that Jupiter's equatorial region
displays a different, harder spectrum than the auroral spots. The
data are best fitted ($\chi^2$/d.o.f. = 26/38) by a two temperature 
{\tt mekal} model, with solar abundances, combined
with a bremsstrahlung continuum and a Gaussian emission line at 1.35
keV (fixed in the fit), corresponding to the energy of the MgXI triplet.
Relevant best fit parameters are listed in Table 2; data, best fit model
and $\chi^2$ contribution from each spectral bin are shown in Fig.~\ref{fig8}.
A fit of the same statistical quality is obtained replacing the 
bremsstrahlung
component with a power law of photon index 1.37 and normalisation at 1 keV
of 1.46 $\times$ 10$^{\rm -6}$ ph cm$^{\rm -2}$ s$^{\rm -1}$ keV$^{\rm -1}$.
The 0.3$-$2.0 keV flux measured for the equatorial region is 4.0 $\times$
10$^{\rm -14}$ erg cm$^{\rm -2}$ s$^{\rm -1}$.

 \begin{table}
      \caption[]{Best fit parameters for the 0.2$-$2~keV spectrum of 
Jupiter's equatorial region (errors are at 90\% confidence)}
         \label{fits_2}
 $
         \begin{array}{lccccc}
            \hline\hline
            \noalign{\smallskip}
 & & \multicolumn{4}{c}{Equator} \\
 & & kT_1^{a} & Norm^{b} & kT_2^{a}
& Norm^{b} \\
            \noalign{\smallskip}
            \hline
            \noalign{\smallskip}
 {\tt mekal}    &  & 0.13^{\rm +0.04}_{\rm -0.02}  &
 11.0^{\rm +5.89}_{\rm -5.18} & 0.44^{\rm +0.06}_{\rm -0.06}
& 9.08^{\rm +1.66}_{\rm -1.54}\\
            \noalign{\smallskip}
 Bremsstrahlung &  & 5^{\rm +136}_{\rm -4} &
 2.07^{\rm +2.34}_{\rm -0.97} & &\\
            \noalign{\smallskip}
            \hline
            \noalign{\smallskip}
Emiss.~line & Energy^{c} & Flux^{d} &
EW^{e} & & \\
            \noalign{\smallskip}
            \hline
            \noalign{\smallskip}
MgXI  & 1.35 & 0.39^{\rm +0.21}_{\rm -0.22} &
144^{\rm +78}_{\rm -81} & & \\
            \noalign{\smallskip}
            \hline\hline
         \end{array}
$\\

$^{a}$ {\tt mekal}/bremsstrahlung temperature in keV\\
$^{b}$ {\tt mekal}/bremsstrahlung normalisation in units of
10$^{\rm -6}$ ph cm$^{\rm -2}$ s$^{\rm -1}$ keV$^{\rm -1}$\\
$^{c}$ Energy of the emission feature in keV (fixed in the 
fits)\\
$^{d}$ Total flux in the line in units of
10$^{\rm -6}$ ph cm$^{\rm -2}$ s$^{\rm -1}$\\
$^{e}$ Line equivalent width in eV\\
   \end{table}

\section{RGS detection}
\begin{figure}
   \centering
   \includegraphics[width=8.8cm]{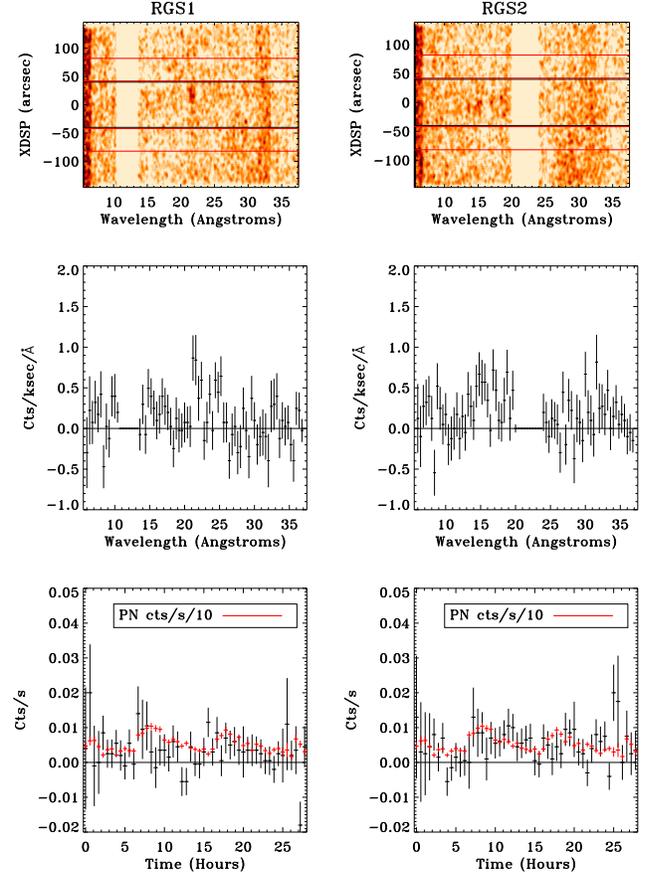}
      \caption{Top: RGS1 (left) and 2 (right) cross-dispersion vs
dispersion plots for the Jupiter observation; the cross-dispersion scale
is referred to the centre of the planet's disk. Middle: RGS1 (left)
and 2 (right) background-subtracted spectra. Bottom: RGS1 (left) and 2 
(right) lightcurves,
with the EPIC-pn lightcurve, appropriately scaled, superposed (in red).}
         \label{fig9}
   \end{figure}

The data from RGS1 and RGS2 (cross-dispersion vs dispersion, before background
subtraction, 1st order only) are shown as colour plots in the top panels 
of Fig.~\ref{fig9}, and the extracted 1st order spectra, background-subtracted,
are in the middle panels. The gross spectrum of the planet is
obtained by integrating along the cross-dispersion direction within 
$\pm$40\arcsec\ of the centre of the planet's disk (i.e. between the two
darker horizontal lines in the top panels of Fig.~\ref{fig9}). The background 
spectrum is the sum of two slices, each 40\arcsec\ wide, centred at 
$\pm$62\arcsec\ from the centre of the planet's disk (i.e. in practice
between the darker and the lighter horizontal lines in Fig.~\ref{fig9}).

It is tantalising to see a bright spot at the expected location of the
OVII triplet (21--22 \AA, or $\sim$0.57 keV) in the RGS1 colour plot, at the
expected location of the North auroral spot (just above the origin in the
cross-dispersion direction, which corresponds to the centre of the planet's
disk). There appears to be an enhancement also at about 15 \AA\ in RGS2,
practically at the centre in the cross-dispersion direction, implying a lower 
latitude on the planet. However, it has to be stressed that these are very
low surface brightness features, which only a longer observation will be
able to confirm.

In order to quantify the possible OVII detection, a box of ~40\arcsec\ in
the cross-dispersion direction and 5 resolution elements ($\sim$0.3 \AA) in 
the dispersion
direction was used to measure the general level of background over CCD4
and the excess at the expected location of OVII: this results in a 2.8$\sigma$
excess in the box centred on OVII. A similar estimate is obtained for the
excess around 15 \AA\ in RGS2.

The bottom panels in Fig.~\ref{fig9} show the background subtracted RGS 
lightcurves with the 0.2 -- 2 keV EPIC-pn lightcurves superposed, appropriately
scaled: the good match of the data from the two instruments adds credibility
to the possible detection of Jupiter X-rays in the RGS.

\section{Discussion and conclusions}

The first {\it XMM-Newton} observation of Jupiter provides interesting new 
information with which we can try and clarify some of the many unresolved 
issues concerning the planet's X-ray emission. Here we consider some of them.

i) The {\it XMM-Newton} EPIC spectra of the North and South auroral regions
of Jupiter can be explained as the superposition of emission lines from
highly ionised oxygen and carbon (the latter in preference to 
sulphur, at least for the North spot). The power law component which is also 
needed to model the spectra could be a pseudo-continuum produced by the 
combination of weaker discrete emission features (although our attempts
at replacing the power law with a combination of emission lines did not
formally improve the quality of the fits). A spectrum consisting 
of many unresolved lines (mostly due to OVII and OVIII ions) was found to 
provide the best fit to the {\it Chandra} data of comet McNaught-Hartley
and was interpreted as the result of electron capture by heavy ions of the 
solar wind colliding with the cometary atmosphere (Kharchenko et al. 
\cite{khar}). A similar interpretation can be proposed for Jupiter's
auroral spots: solar wind ions could be captured and accelerated in the 
planet's magnetic field, and subsequent charge exchange would lead to the 
observed spectra, rich in line emission from highly ionised oxygen and carbon.
The identification of a CVI rather than CV transition (Sect. 4) would also
support an origin from the solar wind, in which carbon is usually fully
ionised.

The North and South spot EPIC spectra (together with the tentative detection
in RGS) clearly show a larger contribution from OVII line emission, relative 
to OVIII; this contrasts with what has been reported from {\it Chandra} 
observations in February 2003
(Elsner et al. 2004, in preparation) and suggests variability, over
a period of a few weeks, in the level of ionisation, and thus of overall 
energy of the ions penetrating Jupiter's upper atmosphere. 

ii) The 0.3$-$2.0 keV fluxes derived from the above spectral analysis
correspond to X-ray luminosities of 0.43, 0.16 and 0.31 GW for
Jupiter's North and South spots, and equatorial region respectively.
Thus the {\it XMM-Newton} results for the auroral regions 
are in good agreement with those of {\it Chandra} two months earlier
(February 2003): at this time Elsner et al. (\cite{Elsner}) measured a 
countrate roughly half of what was observed in December 2000 (Gladstone et al.
\cite{Gladstone02}), when the 0.3$-$2.0 keV
luminosities of North and South spots, and of the equatorial region were
1.0, 0.4 and 2.3 GW respectively. The disk measurement by 
{\it XMM-Newton} appears to be significantly lower than the {\it Chandra} one:
the difference may be due to the use of different extraction regions 
for the disk emission. 


iii) The planet's 10 hr rotational period is observed to modulate the
emission of only the North polar spot, which is found to be at a similar
system III longitude ($\sim$180\degr) as in the {\it Chandra} observations 
of December 2000. If modulation is present in the South spot, 
it must be at a much lower level than observed with {\it Chandra}.

{\it Chandra} images have been used to establish that the North and South 
spots have different morphological appearances, with the North spot well 
localised and the South extended into a band (Elsner et al. \cite{Elsner}).
It is difficult to comment on this with our data, because the {\it XMM-Newton}
spatial resolution is poorer than that of {\it Chandra}. However, 
different morphologies, combined with inclination effects, may be able
to explain the different temporal behaviours at the two poles.

iv) There is no evidence in the {\it XMM-Newton} data for the $\sim$45 min 
oscillations observed with {\it Chandra} in December 2000. This is consistent
with {\it Chandra}'s non-detection in February 2003 and the lack of radio 
quasi-periodic oscillations in {\it Ulysses} data at the same time (Elsner 
et al. 2004, in preparation). We conclude that the oscillations must be a 
transient phenomenon, possibly related to magnetospheric effects on Jupiter, 
which could be variable as a consequence of solar wind variability.
 
v) Jupiter's equatorial X-ray spectrum is harder than that of the polar regions
and resembles what is expected from the scattering of solar X-rays, in
keeping with the prediction of Maurellis et al. (\cite{Maurellis}):
a two-temperature coronal plasma is required to fit the spectrum, as 
well as a contribution from Mg line emission. The latter may be simply
reflected sunlight. Interestingly, a similar feature at $\sim$1.4 keV is
apparent in the {\it XMM-Newton} spectrum of Saturn (Ness et al.
\cite{NessX}). The bremsstrahlung component in Jupiter's spectrum
has an unrealistically high temperature (with very large errors) and is
likely to indicate again the presence of a low level residual flux
which could be due to the superposition of weaker, unresolved lines.

In general Jupiter's equatorial X-ray emission appears to show characteristics 
similar to those observed with {\it XMM-Newton} and {\it Chandra} for Saturn's 
disk emission (Ness et al. \cite{NessX}, \cite{Ness}),
which can also be modelled with a coronal plasma and line emission (likely
to be from oxygen). However, no bright auroral emission is observed 
on Saturn, at least from the South pole (which was the only pole visible
during the {\it Chandra} observation), suggesting different conditions in the
magnetospheres of the two planets.

\begin{acknowledgements}
This work is based on observations obtained with {\it XMM-Newton}, an ESA
science mission with instruments and contributions directly funded by ESA
Member States and the USA (NASA). The MSSL authors acknowledge financial 
support from PPARC.

\end{acknowledgements}

\end{document}